\documentclass[
    ,final            
  ,numberedheadings 
  ]
  {aipproc}

\layoutstyle{8x11double}

\begin{document}

\title{Neutrinos from Supernovas and Supernova Remnants}

\classification{97.60.Bw; 98.58.Mj; 98.70.Sa; 13.15.+g; 14.60.Pq}
\keywords      {supernovas, supernova-remnants, neutrino and $\gamma$ radiation, neutrino oscillations}

\author{M.L. Costantini}{
  address={Universit\'a dell'Aquila and INFN, L'Aquila, Italia}
}

\author{F. Vissani}{
  address={Laboratori Nazionali del Gran Sasso, INFN, Assergi (AQ) Italia}
}

\begin{abstract}
Supernovae (SN) and supernova remnants (SNR) 
have key roles in galaxies, but their 
physical descriptions is still incomplete.
Thus, it is of interest to study neutrino radiation 
to understand SN and SNR better. 
We will discuss:
(1)~The $\sim$10 MeV thermal 
neutrinos that arise from core collapse
SN, that were observed for SN1987A, and 
can be seen with several existing or planned experiments.
(2)~The 10-100 TeV neutrinos expected from 
galactic SNR (in particular from RX~J1713.7-3946) 
targets of future underwater neutrino telescopes.

\end{abstract}

\maketitle
\section{Introduction}
In the economy of a galaxy, the stars with 
mass $\ge 6-10\ M_\odot$ are known to play a prominent role. 
The last instants of their brief life 
(on ten million year scale) is a crucial moment.
In fact, the gravitational collapse of the 
`iron' core leads to supernovae of type II, Ib, Ic and eventually 
to pulsars, neutron stars and stellar black holes. The
gas subsequently expelled by all supernovas including Ia
(that lead to the observed supernova remnants) 
is suspected to be the place where cosmic rays are accelerated.
A complete theoretical understanding of these systems
(and in particular of the supernova)
is not yet available. However, it is known that 
neutrinos of $\sim 10$ MeV are emitted during and just after the 
gravitational collapse,  
and that neutrinos above 1 TeV can be produced
in the supernova remnant at least in special circumstances. Therefore,
neutrinos can be used as diagnostic tools of what happens 
in these interesting and complex systems. With these considerations
in mind, we will discuss here briefly neutrinos from supernova
and from supernova remnants, aiming to outline the main `what', 
`when', `where', and `how' questions.

\paragraph{Outline and references}
In the rest of this section, 
we discuss a number of general 
facts pertaining to supernovas and supernova remnants;
the references that we use are \cite{cavanna,vissani}.
In the second section we concentrate on 
supernova neutrinos; we use \cite{cavanna,vissani}, and 
also \cite{ianni}.
In the third section we describe the calculations 
of~\cite{rxj}. In these 4 papers, one can to find 
more details and a complete list of references.
In the appendix, we show in details a simple way to calculate
neutrinos from cosmic rays, adopted also in~\cite{rxj}.

\paragraph{Generalities}
Before passing to neutrinos, we would like to discuss a few 
general questions.\\
{\bf Guessing `where'}
Let us recall that the only SN within the capability of existing 
neutrino detectors are those 
from our Galaxy (or some irregular galaxy around us).
The best guess we can propose for next galactic 
core collapse SN is
\begin{equation}
{\langle L \rangle=} {10\pm 4.5 \mbox{ kpc}} 
\end{equation}
and it is motivated as follows:\\
1) We are  ${R}=8.5$ kpc from the galactic center.\\
2) The distribution of the 
matter that can go in supernovae is: {$\rho\sim r e^{-r/r_0}$} 
with {$r$}=distance from the center and $r_0= 3$~kpc,
possibly summing a {$\delta(r)$}
to describe the `bar'\\
3) We calculate the 
distribution in function of {$L$}=distance from us,
integrate over the galactic azimuth $\theta$ 
and get the result above.\\
{\bf Guessing `when'}
The rate of core collapse SN in the Milky Way 
is expected to be 
\begin{equation}
{R_{SN}=1/(\mbox{30-70 years})}
\end{equation}
The most reliable method is: count SN in other galaxies, 
and correlate with galactic type.\footnote{The Padova-Asiago 
database \cite{turatto} includes several thousand SN. The Milky Way 
could be Sb or Sb/c, which means a factor 2 uncertainty.}
A similar rate expected for SN Ia.
Possibly, we {missed} several galactic SN due to 
dust.\footnote{Indeed, with about 10 supernovae seen in last 2000 years, 
in order to have a SN each 25 years we need to admit we see just one
SN each 8, that is about the right figure on accounting the presence of 
the dust and the possibility that a supernova explodes on the day sky \cite{clark}.}
For the future, we will have better coverage of galactic SN with neutrinos,  
IR and perhaps with gravitational waves. From 
the absence of neutrino bursts, one derives 
\begin{equation}
{R_{SN}>1/(\mbox{21 years$)$ at 1} \sigma}
\end{equation}
We assume Poisson statistics 
(namely, $\exp(-T R_{SN})=1-C.L$) with $T=24$ years.
To obtain the last number, we note that till 
1986 only Baksan \cite{Baksan} worked with 90 \% DAQ livetime, 
then we assume 100 \% coverage.

We conclude by listing the number of SN
precursors and descendants.
We estimated these populations by 
assuming $R_{SN}^{tot}=1/(\mbox{25 years})$.
\centerline{\begin{tabular}{|c|c|c|}
\hline
Object & Lifetime & Number \\ \hline
Pre-SN with $\nu$ & 20 million $y$ & $400.000$ \\
Pulsars & 2 million $y$ & $40.000$ \\
SNR & $100.000\ y$ &  $2\times 2000$ \\
young SNR & $2000\ y$  & $2\times 40$ \\ \hline
\end{tabular}}
By `Pre-SN with $\nu$' we mean 
core collapse SN. Recall that core collapse SN 
produce neutron stars (NS) or stellar black holes,
and that pulsars  are `active' NS. 
Type Ia SN make white dwarfs, instead, but 
are also supposed to produce SNR (which explains 
the factor 2 above).

\section{$\nu$s from supernovae}
We begin recalling the 
astrophysics of core collapse.\\
The giant stars--{\bf pre-SN}--burn in sequence 
H, He, C and Si, Ne, Mg, Na
{\em etc}, and form an {``onion structure''}, 
with an inert `iron' core.
Violent stellar winds occasionally modify external {\bf mantle}
in latest stages; apparently, this 
happened for SN1987A that was a $\sim 20$ $M_\odot$ {\it blue} giant.
What happens in the {\bf core}?
The gravitational pressure is balanced by  
$e^-$ degeneracy pressure, and the core grows.
As demonstrated by
Chandrasekhar, the pressure of free electrons 
is unable to supply an equilibrium configuration
when $e^-$ become relativistic. Now, 
the iron core mass is $\sim 1.4$ $M_\odot$,
the {\bf collapse} begins and the sequence of the events becomes uncertain
(more on the reference picture, the so-called ``delayed scenario'').
What about the energetic of the collapse?
The total energy of the collapse is {very large},
and 99 \% of this energy is carried away by neutrinos. With
 $M_{ns}/M_{\odot}=1-2$, $R_{ns}=15$~km$(M_{\odot}/M_{ns})^{1/3}$, 
we estimate
$$
{{\cal E}\simeq}\ \frac{3 G_N M^2_{ns}}{7 R_{ns}}=
{(1-5)\times 10^{53}\mbox{ erg}}
$$

\paragraph{The delayed scenario}
This is a pictorial summary of the `delayed scenario' by
Wilson \& Bethe \cite{del}:
\centerline{\includegraphics[width=6.3cm,angle=0]{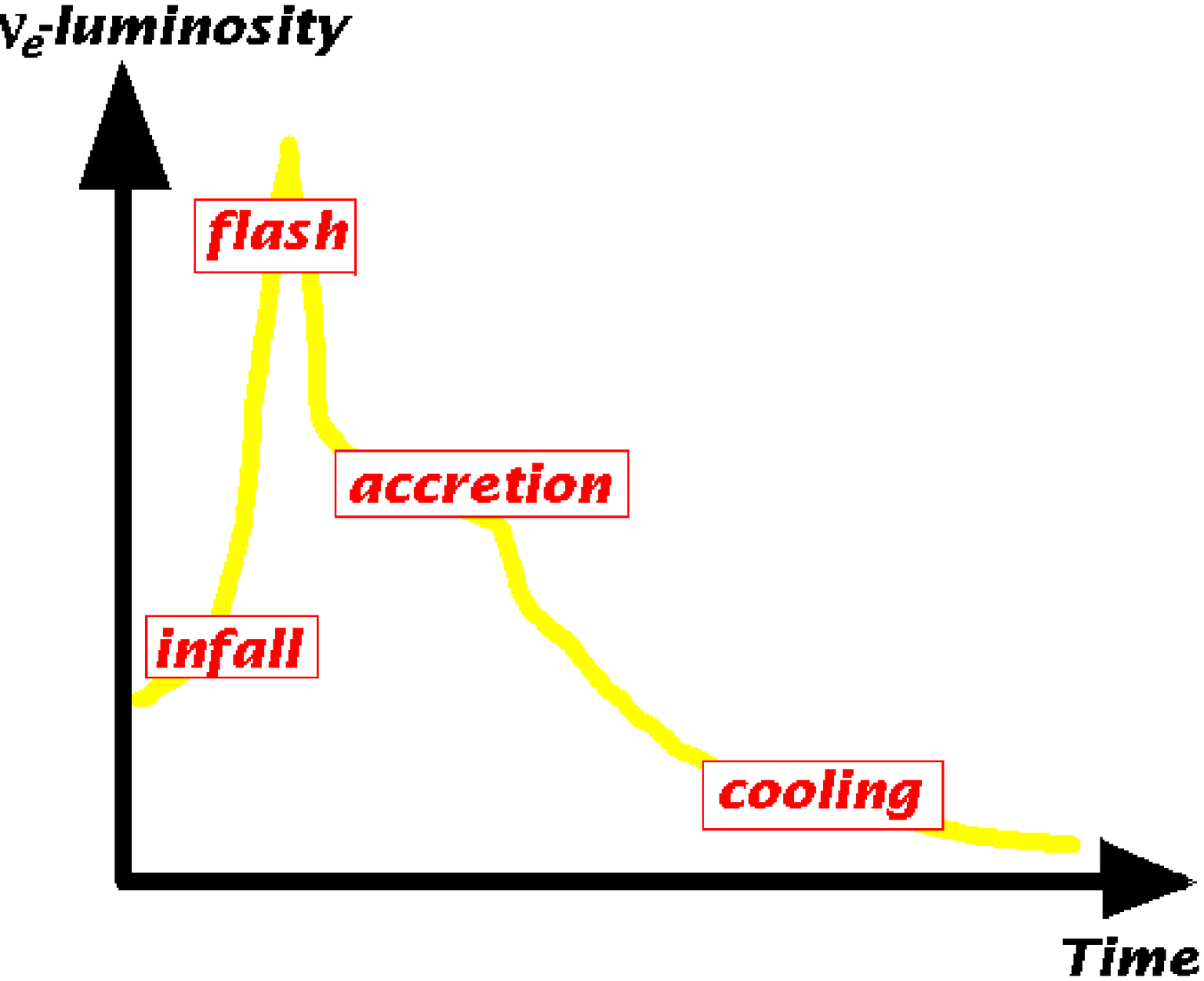}}
\vskip-2mm
The energy radiated in any neutrino
species $e,\bar{e},\mu,\bar{\mu},\tau,\bar{\tau}$
is expected to be the same within a {factor of 
two} \cite{Janka}:
$$
{{\cal E}_e\sim {\cal E}_{\bar e} \sim {\cal E}_x }
$$
where $x$ denotes any among $\mu$, $\bar{\mu}$, $\tau$, $\bar{\tau}$. 
Indeed, in this picture non-electronic neutrinos and antineutrinos are 
produced in a similar amount.

Note that neutrinos are mostly emitted in cooling (80-90 \%) 
and accretion (10-20 \%) phases. The latter one is the phase 
crucial to understand the nature of the 
explosion, whereas the first phase is almost thermal. In this sense,
the ignorance of the detailed mechanism of the explosion does not
preclude a description of the bulk of neutrino radiation (to put it with 
a slogan, ignorance is self-consistent).

Now we come to a prescription for the fluence:
$${
F_i(E)=\frac{{\cal E}_i}{4\pi L^2} \frac{N}{\langle E_i\rangle^2} 
z^\alpha e^{-(\alpha+1) z},}\ \  \ \ \ z=\frac{E}{\langle E_i\rangle}
$$
where ${\langle E_i\rangle}$ is the average energy of the 
neutrino species $i=e,\overline{e},x$;
$N$ ensures that the total energy carried is ${\cal E}_i$.
(If one needs to describe time dependent situations,
${\cal E}_i \to L_i(t)\equiv d{\cal E}_i/dt$,
$\langle E_i\rangle\to \langle E_i(t)\rangle$, 
$\alpha\to \alpha(t)$.)
The expectations for time integrated quantities are:
$$
\begin{array}{ll}
{\langle E_{\bar e}\rangle}=12-18\mbox{ MeV}, &
{\langle E_{x}\rangle}/{\langle E_{\bar e}\rangle}=1-1.2\\
{\cal E}_{\bar e}=(2-10)\times 10^{52}\mbox{ erg} & 
{\cal E}_{x}/{\cal E}_{\bar e}=1/2-2
\end{array}
$$
One guesses ${\cal E}_{e}={\cal E}_{\bar e}$
(not so important); the 
$\nu_e$ average energy can be estimated from the emitted 
lepton number.

\paragraph{Oscillations of supernova neutrinos}
Oscillations of 
SN neutrinos are pretty simple to describe.
To account for oscillations we need to 
assign just 2 functions, ${P_{ee}}$ and 
${P_{\bar{e}\bar{e}}}$:
$$
\begin{array}{l}
{\bullet} 
F_e\begin{array}[t]{l} 
=F_e^0\ P_{ee}+F_\mu^0\ P_{\mu e}+F_\tau^0\ P_{\tau e}\\
=F_e^0\ P_{ee}+F_x^0\ (1-P_{ee})
\end{array}\\
{\bullet} 
F_e +F_\mu +F_\tau =F_e^0 +F_\mu^0 +F_\tau^0 
\end{array}
$$
and similarly for antineutrinos
(we will consider only oscillations of massive $\nu$s).
The relevant densities to calculate $P_{ee}$ and $P_{\bar{e}\bar{e}}$
are $\rho_{sol}\sim 10$~gr/cc (He) and $\rho_{atm}\sim 10^3$~gr/cc (C+O).

\begin{table}[t]
\begin{tabular}{llll}\hline
Emission & Good& Bad & Remarks \\ \hline
cooling & strong $\nu$ radiation & uncertainties, small effect & $F_x^0\sim F_e^0$ \\ \hline
accretion & strong $\nu$  radiation & uncertainties! & $F_x^0\sim F_e^0/2$ \\
\hline
neutronization & clean  signal & weak $\nu$ radiation & $F_x^0\sim 0$ \\
\hline
neutroniz.++  with rotation & clean and strong sign. & uncertain!!  & $F_x^0\sim 0$ {\it (LSD?)}\\
\hline
\end{tabular}
\caption{Depending on the phase of neutrino emission, it is possible 
to learn more on $U_{e3}$ studying of SN neutrinos. 
In the table, goods and bads of each specific phase are 
mentioned. (LSD refers to the observations of 
the Mont Blanc detector, that according to Imshennik and Ryazhskaya \cite{2stageIR}
could be interpreted assuming an intense neutronization phase 
before the main $\nu$ burst.)}
\label{tab1}
\end{table}
There is a great interest on  the unknown parameter
$U_{e3}$, and supernovas offer a chance to learn more on that.
However, we should keep in mind the uncertainties 
mentioned above. Let us try to discuss this point further.
We begin with the formula $F_e=F_x^0-P_{ee} (F_x^0-F_e^0)$.
If we have normal mass hierarchy:
$$
P_{ee}=\left\{
\begin{array}{ll}
\sin^2\theta_{13}\sim 0 & \theta_{13}\mbox{ `large', }>1^\circ \\
\sin^2\theta_{12}\sim 0.3 & \theta_{13}\mbox{ `small', }<0.1^\circ 
\end{array}
\right.
$$
Now  we can 
ask a precise question on $U_{e3}$:
\begin{quotation}
{\em Can we distinguish the two cases?}
\end{quotation}
Table \ref{tab1} provides a check-list.
Note that if the mantle is stripped off till 
densities $\rho> 10$~gr/cc (e.g., with SN~Ic)
we have vacuum oscillations, $0.3\to 0.6$ (Selvi). 
This is rare, but not impossible.

\begin{figure}[t]
\includegraphics[width=7cm,angle=270]{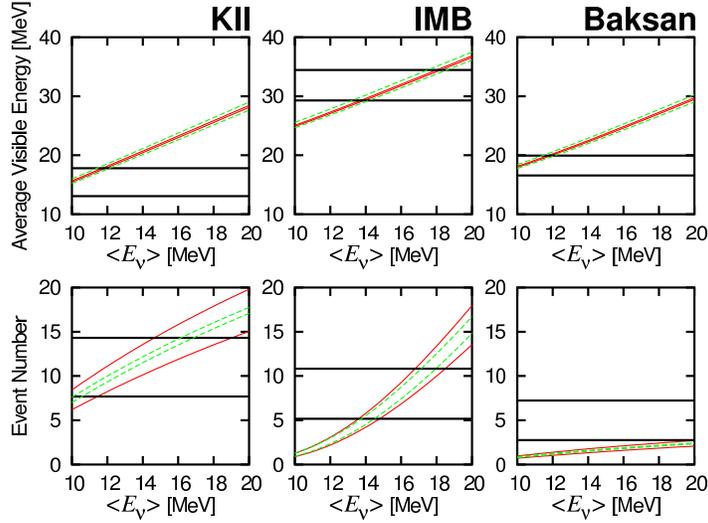}
\caption{Average energies and number of events for
three detectors. The horizontal lines describe the measured values,
the oblique lines describe the theoretical expectations.}
\label{fig1}
\end{figure}

\paragraph{SN1987A}
The first detection of SN neutrino is of epochal importance.
These observations
fit into the `standard' picture for neutrino emission 
(see next figure), but there are some 
\underline{puzzling} aspects:\\
1.\ IMB \cite{IMBdata} and Kamiokande-II
\cite{KIIpaper} find forward peaked distributions;
e.g., $\langle \cos\theta\rangle $ are ${\sim 2\ \sigma}$ 
above expectations.\\
2.\ $\langle E_{vis}^{\rm KII}\rangle\sim {15}$~MeV and
 $\langle E_{vis}^{\rm IMB}\rangle\sim {30}$~MeV
($\pm 2.5$~MeV)  
are quite different even correcting for efficiencies.\\
3.\ {The time sequence} of events looks different;
when combined  not so bad (but abs.\ time is unknown).\\
4.\ The {5 LSD} events, occurred 4.5 hours before the main signal,
cannot be accounted for.\\
In figure \ref{fig1}, 
will stress the {consistency} of the `standard' interpretation, but 
the space for non-standard ones is not small 
(not only due to limited statistics).
A reasonable agreement with expectations is obtained if:
$$
\begin{array}{l}
{\langle E_{\bar e}\rangle\equiv E_0=12-16\mbox{ MeV}} \\
{{\cal E}=(2-3)\times 10^{53}~\mbox{erg}}\\
\mbox{{zero or a few} $\nu_e\ e\to\nu_e\ e$ events in KII}
\end{array}
$$

\section{$\nu$s from SNR}
The leftover gas is the SNR.
The kinetic energy is a few times $10^{51}$ erg,
which means that the gas is in free expansion with
velocities of about $\sim 4000$ km/sec.
There are various phases of the SNR, according to their age,
and various shapes (shell, plerionic, or mixed). 

An argument by Ginzburg and Syrovatskii suggests SNR as 
main source of galactic cosmic rays (CR):\\
1) The Milky Way irradiates CR. 
Take $V_{CR}=\pi R^2 H$ with $R\sim 15$ kpc, $H\sim 5$ kpc 
as the volume
of confinement. Take $\tau_{CR}=5\times 10^7$ 
years as CR lifetime in the Galaxy. We get the CR luminosity:
$$
{\cal L}_{CR}=\frac{ V_{CR}\cdot \rho_{CR}}{\tau_{CR}}=
0.9\times 10^{41} \mbox{erg/s}
$$
2) We have a new SN each $\tau_{SN}\sim 25$ year, with  
about ${\cal E}\sim 10^{51}$ erg in kinetic energy, that is
$$
{\cal L}_{SN}=\frac{{\cal E}}{\tau_{SN}} =
1.2\times 10^{42} \mbox{erg/s}
$$
Comparing the two formulae, we see that if a SNR is able to convert a 
fraction $f_{CR}\sim 5-10$~\% of the injected energy into CR, we are home
(the numbers quoted above shouldn't be taken too seriously, but
this 40-years-old argument maintains its appeal).

We proceed describing the system.
In 2000 years, the SNR proceeds by $\sim 10$ pc. The density
is $\sim 0.2$ protons/cm$^3$,  
too faint to permit a significant production of secondaries.
In the galaxy, the  most dense  
non-collapsed objects are the molecular clouds, whose
density can reach $10^4$ protons/cm$^3$.
The 2 objects form:
\centerline{\includegraphics[width=4.5cm]{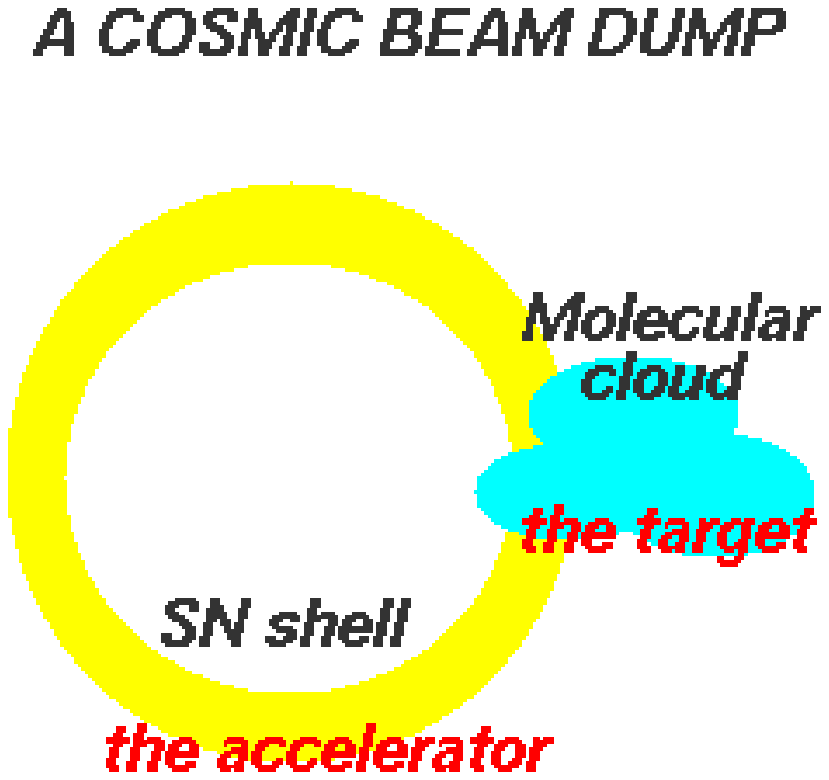}}
That is an ideal configuration, since
$$
\mbox{ CR collisions}\to 
\left\{
\begin{array}{ll}
\pi^0\to & \mbox{ high energy }\gamma\\
\pi^\pm\to & \mbox{ high energy }\nu_\mu,\nu_e
\end{array}
\right.
$$
Thus, we expect to have $\gamma$ {\it and} neutrinos.

There is some evidence that 
the system called RX J1713.7-3946 is a `cosmic beam dump'. Indeed,\\
1) It is seen in X-rays, with many details\\
2) It is in Chinese Annales, 393 A.D. \cite{wang} \\
3) A molecular cloud seen in CO and 21 cm H line \\
4) And most interestingly, CANGAROO (since 2000, see \cite{CA1}) 
and H.E.S.S.\ (since 2004, see \cite{hess}) do see TeV $\gamma$ rays.\\
The distance is $L=1$ kpc, the angular size about 1$^\circ$,
the density of the cloud $\sim 100$ part/cm$^3$. The source is 
transparent to gamma rays, so that the neutrino flux can be calculated 
easily and reliably as shown in the appendix. (However, 
no item above seems to be waterproof, what 
we discuss is just an appealing 
interpretation.)

We assume that the cosmic ray spectrum:
$$
{F_p}=K E^{-\Gamma},\ \ \ 
\Gamma=2-2.4
$$
interacts with the molecular cloud and is the main source of 
visible gamma rays, through the reaction $p\to \pi^0\to\gamma$. 
Neutrinos originate from similar processes, e.g., 
the reaction $p\to \pi^+\to\mu^+\to \nu_e$ will yield $\nu_e$.
Using the flux  $F_\Gamma=1.7\cdot 10^{-11} (E/\mbox{TeV})^{-2.2} \ \rm TeV^{-1} cm^{-2} s^{-1} $
measured by H.E.S.S.\ in the range between 1-40 TeV we get:
$$
\begin{array}{l}
F_{\nu_\mu}^0=7.3\times 10^{-12}\ 
\left(\frac{E}{\rm TeV}\right)^{-2.2}
\frac{1}{\rm TeV cm^{2} s }\\[1ex]
F_{\overline{\nu}_\mu}^0=7.4\times 10^{-12}\ 
\left(\frac{E}{\rm TeV}\right)^{-2.2}
\frac{1}{\rm TeV cm^{2} s }\\[1ex]
F_{\nu_e}^0=4.7\times 10^{-12}\ 
\left(\frac{E}{\rm TeV}\right)^{-2.2}
\frac{1}{\rm TeV cm^{2} s }\\[1ex]
F_{\overline{\nu}_e}^0=3.0\times 10^{-12}\ 
\left(\frac{E}{\rm TeV}\right)^{-2.2}
\frac{1}{\rm TeV cm^{2} s }
\label{flussi0}
\end{array}
$$
3 flavor oscillations take the simplest form (vacuum averaged or 
Gribov-Pontecorvo) and can be included easily. This leads to 
an important modification of the fluxes.

Finally we pass to the
signals of neutrinos.
$\nu$ interactions are due to deep elastic scattering.
The simplest and most traditional observable is  
{induced muons}, that can be correlated 
to the source by mean of an angular cut.
It is important to recall that high energy $\nu_\mu$ are to some extent 
{absorbed} from the Earth, and 
that when the source is above the horizon, it is impossible 
to see anything due to the {background} from atmospheric $\mu$.
Along with {oscillations}, these effect decrease the observable 
neutrino signal.
For an ideal detector, located in the Mediterranean, with 
$\mbox{Area=1 km}^2$, 
$\mbox{Data taking=1 year}$,
$E_{thr.}=50\mbox{ GeV}$
we find that the number of expected events 
is {about 10}\\ (this was 30 if 
oscillations, absorption and $\mu$-background were ignored, or
even 40 if the slope was $\Gamma=2.0$).

\section{Discussion and Perspectives}
We do not have a clear understanding of how SN
explode. Perhaps, because it is a very difficult problem, 
or there is some missing ingredient,
perhaps the answer is not unique (a combination of various mechanisms?),
... the confusion could persist even after next galactic SN.

$\nu$s from next galactic SN have an impressive potential
to orient our understanding. The hypothesis of an ``accretion phase'',
implying 10-20 \% of the emitted energy, 
can certainly be tested. 
SN1987A does not contradict the `delayed scenario' seriously 
(but does not help much either).
There are chances to learn 
on oscillations and in general on particle physics.
The possibility to use $\nu_e$ and 
neutral current events deserves consideration.

$\nu$s from SNR are an uncharted territory. 
Recent results from H.E.S.S.\ 
motivated us to consider one specific SNR (but we expect new results
and surprises with $\gamma$ rays). 
Sure enough, CR acceleration in SNR is not fully understood,
and there are other possible sources for TeV $\nu$ astronomy,
however the number of events we found (about 10/km$^2$ y) suggests
the need of large exposures. It is important to
improve our theoretical tools to describe SNR $\gamma$ and $\nu$.

\vskip2mm
\noindent{\bf APPENDIX: Cosmic rays and neutrinos.}
We report here on the details of the calculation of (anti) neutrino 
spectra from the decays of secondary particles produced in pp
interaction in astrophysical beam-dumps (table 1 of \cite{rxj}). 
As a first approximation, we add the contributions
to $\nu$-emission from two chains: (1) $ p \to m^{\pm} \to (anti)\nu $ and (2) 
$ p \to m^{\pm} \to \mu^{\pm} \to (anti)\nu $,
where $m^{\pm}= \{ \pi^{\pm},K^{\pm} \}, 
\nu = \{ \nu_e, \bar{\nu}_e, \nu_{\mu}, \bar{\nu}_{\mu} \} $. \\ 
If we assume scaling, and take a power-law proton spectrum  
$F_p(E_{p})=K  E^{-\Gamma}$, we can express the (anti) 
neutrino spectra
from (1) and (2) as:
\begin{equation}
\label{psinu}
F^{(1),(2)}_{\nu}(E_{\nu})=
\frac{\Delta X}{\lambda_p}\cdot \psi^{(1),(2)}_{\nu} \cdot F_p(E_{\nu}) 
\end{equation}
where $\psi_{\nu}$ indicate the neutrino emissivity coefficient, 
defined in \cite{bere79},
$\Delta X$ the column density traversed by the protons 
and $\lambda_p$ the interaction length of CR.
Similarly, the $\gamma$-spectrum from $p\to \pi^0\to \gamma \ $ is:
\begin{equation}
\label{psigamma}
F_{\gamma}(E_{\gamma})=
\frac{\Delta X}{\lambda_p}\cdot \psi_{\gamma} \cdot F_p(E_{\gamma}),
\mbox{ where }
\psi_{\gamma}=\frac{2 Z_{p \pi_0}}{\Gamma} 
\end{equation}
In equations \ref{psinu} and \ref{psigamma} we assume a thin target.
When we assume scaling, following Gaisser (see \cite{gai}, 
formulas 4.1 and 4.2 
in approximation 3.31) we have:
\begin{equation}
F^{(1)}_{\nu_{\mu}}=\Delta X  \sum_{m^{+}=\pi^+,K^+} 
 \int_{\frac{E_{\nu_{\mu}}}{1-r_{m}}}^{\infty}
\frac{dn_{\nu_{\mu} m^{+}}}{dE_{\nu_{\mu}}} \ D_{m^{+}}(E_{m^{+}}) \ dE_{m^{+}}
\end{equation}
with:
\begin{equation}
\frac{dn_{\nu_{\mu}m^+}}{dE_{\nu_{\mu}}}=\frac{B_{m^+ }}{1-
r_{m}} \ \frac{1}{E_{m^+}},\ 
D_{m^+}=\frac{Z_{pm^+}}{\lambda_p} F_p(E_{m^+})
\end{equation}
where $B_{m^+}$ is the branching ratio for 
$m^+ \to \mu^+ \nu_{\mu}$, $r_{m}=m^2_{\mu}/m^2_{m}$
(subscript $m=\{\pi,K \}$),
and $Z_{pm^+}$ are the spectrum-weighted momenta, defined in \cite{gai}.
We recall that in the scaling hypothesis the spectrum-weighted 
momenta depend 
only from the spectral index $\Gamma$.
We can get the $\bar{\nu}_{\mu}$ spectrum from the 
same formula, replacing  $m^+$ with $m^-$.
Evaluating the integrals, we get:
\begin{equation}
F^{(1)}_{\nu_{\mu}}=\frac{\Delta X}{\lambda_p} \ 
F_p(E_{\nu_{\mu}}) \sum_{m^{+}=\pi^+,K^+}  B_{m^+}
\frac{Z_{pm^+}}{\Gamma } (1-r_{m})^{\Gamma-1}
\end{equation}
and so:
\begin{equation}
\psi^{(1)}_{\nu_{\mu}}=\sum_{m^{+}=\pi^+,K^+}  B_{m^+}
\frac{Z_{pm^+}}{\Gamma } (1-r_{m})^{\Gamma-1}
\end{equation}
Similarly, for $ \psi^{(2)}_{\nu_{\mu}} $ we have 
(see \cite{gai}, formula (7.14), in which we 
replace $\phi_{\pi}$ with
the factor $[ \Delta X  Z_{p m^-} 
  B_{m^-}   F_p(E_{\nu_{\mu}}) ] / \lambda_p $,
and then divide by $ \Delta X F_p / \lambda_p$):
\begin{eqnarray}
\label{psi2}
& \quad &   \psi^{(2)}_{\nu_{\mu}} =
 \sum_{m^{-}=\pi^-,K^-} Z_{p m^-} 
B_{m^-} \frac{1-r_{m}^{\Gamma}}{\Gamma (1-r_{m})}
\Bigg\{ \langle y_{0\nu_{\mu}}^{\Gamma-1} \rangle + \nonumber \\
   &\quad &     + \Bigg[ 1+r_{m} - \Bigg( \frac{2 \Gamma r_{m}}{1-r_{m}^\Gamma} \Bigg)
   \ \frac{1-r_{m}^{\Gamma-1}}{\Gamma-1} \Bigg]   \frac{\langle  
y_{1 \nu_{\mu}}^{\Gamma-1} 
\rangle }{1-r_{m}}  \Bigg\}
\end{eqnarray}
where the moments:
\begin{equation}
\langle y^{\Gamma-1}_{i \nu}\rangle= \int_{0}^{1} y^{\Gamma-1} g_{i \nu}(y) dy
\end{equation}
for (anti)$\nu_{e}$ and (anti)$\nu_{\mu}$ are given in 
table 7.3 from \cite{gai} (the $g_{i\nu}(y)$ 
are given in table 7.2 from \cite{gai}).
We can obtain $ \psi^{(2)}_{\bar{\nu}_{\mu}}$ from eq.~\ref{psi2}, 
replacing  $Z_{p m^-}$ with  $Z_{p m^+}$.
In our approximation, the (anti)$\nu_e$ emissivity coefficients 
are given by eq.~\ref{psi2}, 
inserting respectively
$Z_{p m^+}$ for $ \psi_{\nu_{e}}$, and  $Z_{p m^-}$ for $ \psi_{\bar{\nu}_{e}}$
and the appropriated moments.
The numerical values of $ k_{\nu}=(\psi^{(1)}_{\nu}+\psi^{(2)}_{\nu})/\psi_{\gamma}$ for several 
values of $ \Gamma$, obtained using the 
spectrum-weighted momenta
from figure 5.5 by \cite{gai}, are reported in 
table 1 by \cite{rxj}.\footnote{In 
our calculation, we omitted the contribution of $\nu$'s produced in 
other decay modes of $\pi$ and $K$,
as well as in $K^0$'s decay chains, and so we introduce and error 
at several \% level.
In the calculation made in \cite{bere79}, those contributions are 
taken into account.
Moreover, the values of $Z_{pm}$ used in \cite{bere79}
differ from our values at several \% level.   
We note that in the calculation made in \cite{bere79}, only
the first addend in the parenthesis, $ \langle y_0^{\Gamma-1} \rangle $, 
is taken into account, that introduces an error at the few \% level.}

\end{document}